\documentclass[prl,twocolumn,showpacs,comment]{revtex4}
\usepackage{amsmath,amsthm,amsfonts,amscd,mathrsfs,pifont,bm}
\usepackage{eucal}

\renewcommand{\atop}[2]{\genfrac{}{}{0pt}{}{#1}{#2}}

\begin{document}

\title  {Are Bosonic Replicas Faulty?}

\author {Vladimir Al.~Osipov} \email
{vosipov@hit.ac.il}
\author {Eugene Kanzieper} \email
{eugene.kanzieper@weizmann.ac.il}

\affiliation
        {Department of Applied Mathematics, H.I.T.---Holon Institute of Technology
        Holon 58102, Israel}
\date   {April 23, 2007}

\begin  {abstract}
Motivated by the ongoing discussion about a seeming asymmetry in the
performance of fermionic and bosonic replicas, we present an exact,
nonperturbative approach to {\it both} fermionic and bosonic
zero-dimensional replica field theories belonging to the broadly
interpreted $\beta=2$ Dyson symmetry class. We then utilise the
formalism developed to demonstrate that the bosonic replicas {\it
do} correctly reproduce the microscopic spectral density in the 
QCD-inspired chiral Gaussian unitary ensemble. This disproves the myth
that the bosonic replica field theories are intrinsically faulty.
\end{abstract}

 \maketitle

{\it Introduction.}---Since the mid-1990s, there has been a revived
interest in the field theoretic approaches tailor-made to the
analysis of {\it interacting} disordered and quantum chaotic
systems. In particular, the exact Keldysh \cite{KA-1999} and
approximate supersymmetry \cite{SE-2005} techniques have been
conceived to offer a nonperturbative alternative to the notoriously
known {\it replica field theories} \cite{W-1979,ELK-1980,F-1983}
whose legitimacy has been questioned \cite{VZ-1985} for more than
two decades. Sadly, the newly proposed field theoretic approaches
\cite{KA-1999,SE-2005} have not yet evolved into efficient
calculational tools, and their success \cite{AK-2000} has been very
limited.

At the same time, substantial progress \cite{K-2002,K-2005,SV-2003}
was achieved over the past few years in resolving controversies
surrounding nonlinear replica $\sigma$ models. Specifically, the
{\it fermionic} version \cite{ELK-1980,KM-1999} of a replica field
theory considered in the so-called zero-dimensional
(random-matrix-theory \cite{M-2004}) limit was proven \cite{K-2002}
to be exactly integrable. This observation brought into play the
whole machinery of the theory of integrable hierarchies
\cite{DKJM-1983} and eventually resulted in reconstructing
\cite{K-2002,K-2005} the exact spectral densities and/or correlation
functions for the paradigmatic Gaussian unitary ensemble (GUE), the
QCD inspired chiral GUE (chGUE) \cite{VZ-1993} and Ginibre's
ensemble \cite{G-1965} of complex non-Hermitean random matrices. The
exact fermionic replicas also have immediate implications for the
nonperturbative physics of the 1D lattice impenetrable bosons
\cite{GS-2006}. The supersymmetric variation \cite{SV-2003} of exact
replicas \cite{K-2002} has already produced important new results
\cite{SV-2004} for the QCD at nonzero chemical potential.

The present Letter, prompted by the ongoing discussion
\cite{VZ-1985,DV-2001,NK-2002,F-2002} about a seeming asymmetry in
the performance of fermionic and bosonic replicas \cite{Rem-Bos}
(fermionic-bosonic dichotomy), addresses the problem of
integrability of zero-dimensional {\it bosonic} replica field
theories much in line with the ideas of Refs. \cite{K-2002,K-2005}.
Having formulated a general nonperturbative theory of {\it both}
fermionic and bosonic replicas, we further concentrate on the chGUE
matrix model and develop an integrable theory of the corresponding
nonlinear {\it bosonic} replica $\sigma$ model. Contrary to the
claims made in the literature \cite{DV-2001}, the latter is shown to
produce the {\it exact} expression for the chGUE density of
eigenlevels in the physically relevant limit of infinite-dimensional
matrices. This achievement, representing the main outcome of our
study, provides strong evidence that {\it the bosonic replicas are
as good and reliable as the fermionic ones}. We conjecture that the
above statement holds in the whole generality, no matter what
particular random matrix model is being treated.

{\it How replicas arise and why they are tricky.}---Replica field
theories (be it the original bosonic formulation invented by Wegner
\cite{W-1979} or its fermionic counterpart \cite{ELK-1980} further
extended by Finkelstein \cite{F-1983} to accommodate the interaction
effects) are based on the identity
\begin{eqnarray} \label{RL}
    \log \, Z = \lim_{n\rightarrow \pm 0} \frac{Z^n-1}{n}
\end{eqnarray}
which can be very useful \cite{EA-1975} in evaluating the average
$\langle \log Z \rangle$. Upon assigning $Z$ the meaning of a
quantum partition function $Z({\boldsymbol \varsigma}) =
\prod_{\alpha=1}^p \det(\varsigma_\alpha- {\cal H})$ of a system
characterised by a stochastic Hamiltonian ${\cal H}$, the identity
(\ref{RL}) can be utilised to represent the average $p$-point Green
function $ G({\boldsymbol \varsigma})
    =
    \langle
    \prod_{\alpha=1}^p \, {\rm Tr}\, \left(\varsigma_\alpha - {\cal H}
    \right)^{-1}
    \rangle
$ in terms of the average characteristic polynomials
\begin  {eqnarray}
        \label{annealed}
        Z_n({\boldsymbol \varsigma}) =
        \left<
        \prod_{\alpha=1}^p
        \det{}^n (\varsigma_\alpha - {\cal H}) \right>,\;\;\;
        {\boldsymbol \varsigma} = (\varsigma_1,\cdots,\varsigma_p),
\end    {eqnarray}
to be referred to as the replica partition function. Notice that
Eq.~(\ref{annealed}) is defined for $n\in {\mathbb R}$
\cite{Rem-phys}. The recipe, known as the replica limit, reads
\begin{equation}
\label{p-cf}
    G({\boldsymbol \varsigma})
    =
    \lim_{n\rightarrow \pm 0} \, \frac{1}{n^p}\,
    \partial_{\varsigma_1} \cdots  \partial_{\varsigma_p}
    \,
    Z_n ({\boldsymbol \varsigma}).
\end{equation}
Equation (\ref{p-cf}) assumes a mutual commutativity of the
following operations: the replica limit, differentiation, disorder
averaging denoted by the angular brackets $\langle \cdots \rangle$,
and a thermodynamic limit, if necessary.

Seemingly innocent at first glance, the prescription (\ref{p-cf}) is
much trickier than one could na\"{i}vely expect. Indeed, in order to
calculate the replica partition function $Z_n({\boldsymbol
\varsigma})$ nonperturbatively, a field theorist interprets
$Z^n({\boldsymbol \varsigma})$ in Eq.~(\ref{annealed}) as a
substitute for $|n| \in {\mathbb Z^+}$ identical noninteracting
copies, or replicas, of the original random system. Each copy,
exemplified by the product $\prod_{\alpha=1}^p\det(\varsigma_\alpha
- {\cal H})$ of $p\ge 1$ single determinants, is represented by a
functional integral over an auxiliary field which is either
fermionic or bosonic by nature, depending on the sign of $n$.
Exponentiating a random Hamiltonian ${\cal H}$, such a
representation facilitates a nonperturbative averaging over the ensemble
of stochastic Hamiltonians in Eq.~(\ref{annealed}) and eventually
results in effective field theories defined on either a compact
\cite{ELK-1980} (fermionic, $n \in {\mathbb Z}^+$) or a noncompact
\cite{W-1979} (bosonic, $n \in {\mathbb Z}^-$) manifold. Such a
replica mapping, $Z_{n \in {\mathbb R}}({\boldsymbol \varsigma})
    \stackrel{{\rm map}}{\longrightarrow}
    {\tilde Z}_{n \in {\mathbb Z}^\pm}({\boldsymbol \varsigma})$,
clearly indicates the key problem of replicas. By derivation, the
validity of ${\tilde Z}_{n \in {\mathbb Z}^\pm}({\boldsymbol
\varsigma})$ is restricted to $n \in {\mathbb Z}^\pm$, which is not
enough for implementing the replica limit (\ref{p-cf}) determined by
the behaviour of ${\tilde Z}_n({\boldsymbol \varsigma})$ in the {\it
vicinity} of $n=0$. This mismatch between the ``available'' and the
``needed'' is at the heart of the trickery with which the replica field
theories are often charged \cite{VZ-1985}.

The canonical way to bridge this gap is to determine ${\tilde
Z}_n({\boldsymbol \varsigma})$ for $n \in {\mathbb Z}^\pm$, and then
attempt to analytically continue ${\tilde Z}_n({\boldsymbol
\varsigma})$ away from $n$ integers, in general, and to a proper
vicinity of $n=0$, in particular. Since performing an analytic
continuation based on an {\it approximate} result is a
mathematically questionable procedure, the evaluation of ${\tilde
Z}_n({\boldsymbol \varsigma})$ must be done {\it exactly}. Below, we
will show how such a nonperturbative calculation can be carried out
in quite a general setting. The approach to be presented applies to
the matrix models belonging to the broadly interpreted Dyson's
$\beta=2$ symmetry class \cite{M-2004,AZ-1997} and is by far more
flexible and efficient than the one of Ref. \cite{K-2002}.

{\it Nonperturbative approach to replicas.}---Let us concentrate on
the fermionic and/or bosonic replica field theories whose mapped
partition functions admit the eigenvalue representation ($n$ is
supposed to be positive)
\begin{equation}
\label{rpf-1}
    {\tilde Z}_{n}^{\rm(f/b)}({\bm {\varsigma}}) = \int_{{\cal D}^{n}} \prod_{k=1}^{n}
        d\lambda_k\, \Gamma({\boldsymbol \varsigma};\lambda_k)\, e^{-V_n(\lambda_k)}
    \cdot
    \Delta_{n}^2({\boldsymbol \lambda}).
\end{equation}
Here $V_n(\lambda)$ is a ``confinement potential'' which may depend
on the replica index $\pm n$; $\Gamma({\boldsymbol
\varsigma};\lambda)$ is a function accommodating relevant physical
parameters ${\boldsymbol \varsigma}$ of the theory [they are not
necessarily the energies specified in Eq.~(\ref{annealed})]. To
treat the fermionic and bosonic replicas on the same footing, the
integration domain ${\cal D}$ was chosen to be \cite{Int-D} ${\cal
D}=\bigcup_{j=1}^r [c_{2j-1},c_{2j}]$.

To detemine the replica partition function ${\tilde
Z}_{n}^{\rm(f/b)}({\bm {\varsigma}})$ nonperturbatively, we adopt
the ``deform-and-study'' approach, a standard string theory method
of revealing hidden structures. Its main idea consists of
``embedding'' ${\tilde Z}_{n}^{\rm(f/b)}({\bm {\varsigma}})$ into a
more general theory of $\tau$ functions
\begin{eqnarray}
\label{tau-f}
    \tau_{n}^{(s)}({\boldsymbol \varsigma}; {\boldsymbol t}) &=& \frac{1}{n!}
    \int_{{\mathcal D}^{n}} \prod_{k=1}^{n}
        d\lambda_k\,  \Gamma({\boldsymbol \varsigma};\lambda_k) \nonumber \\
        &\times& e^{-V_{n- s}(\lambda_k)} \,
        e^{v({\boldsymbol t};\lambda_k)} \cdot
     \Delta_{n}^2({\boldsymbol \lambda})
\end{eqnarray}
which posses the infinite-dimensional parameter space\linebreak
${\bm t}=(t_1, t_2,\cdots)$ arising as the result of the
${\boldsymbol t}$-deformation $v({\boldsymbol t};\lambda) =
\sum_{j=1}^\infty t_j \lambda^j$; the auxiliary parameter $s$ is
assumed to be an integer, $s \in {\mathbb Z}$. Studying the
evolution of $\tau$ functions in the extended $(n,s,{\bm t}, {\bm
\varsigma})$ space allows us to identify the highly nontrivial,
nonlinear differential hierarchical relations between them.
Miraculously, a projection of these relations, taken at $s=0$, onto
the hyperplane ${\bm t}={\bm 0}$,
\begin{eqnarray}
\label{proj}
    {\tilde Z}_n^{\rm (f/b)}({\bm\varsigma}) = n!\,\tau_n^{(s)}({\bm \varsigma}; {\bm t})
    \Big|_{\atop{{\bm t}={\bm 0}}{s=0}},
\end{eqnarray}
generates, among others, a closed nonlinear differential equation
for the replica partition function ${\tilde Z}_n^{\rm (f/b)}({\bm
{\varsigma}})$. Since this {\it nonperturbative} equation appears to
contain the replica (or hierarchy) index $n$ as a parameter, it is
expected \cite{K-2002} to serve as a proper starting point for
building a consistent analytic continuation of ${\tilde Z}_n^{\rm
(f/b)}({\bm {\varsigma}})$ away from $n$ integers.

Having formulated the crux of the method, let us turn to its
detailed exposition. The two key ingredients of the exact theory of
$\tau$ functions are (i) the bilinear identity \cite{DKJM-1983} and
(ii) the (linear) Virasoro constraints \cite{MM-1990}.

(i) The bilinear identity encodes an infinite set of hierarchically
structured nonlinear differential equations in the variables
$\{t_j\}$. For the model introduced in Eq.~(\ref{tau-f}), the
bilinear identity reads \cite{TSY-1996,OK-2007}:
\begin{widetext} \vspace{-0.5cm}
\begin{equation} \label{bi-id}
    \oint_{{\cal C}_\infty} dz \,e^{a\, v(\bm{t-t^\prime};z)}\left(
    \tau_{n}^{(s)}(\bm{t}-[\bm{z}^{-1}])\,
    \frac{\tau_{m+1}^{(m+1+s-n)}(\bm{t^\prime}+[\bm{z}^{-1}])}{z^{m+1-n}}\, e^{v(\bm{t-t^\prime};z)}
    -\tau_m^{(m+s-n)}(\bm{t^\prime}-[\bm{z}^{-1}]) \frac{\tau_{n+1}^{(s+1)}
    (\bm{t}+[\bm{z}^{-1}])}{z^{n+1-m}}\,
    \right)=0.
\end{equation}
\end{widetext}
Here, $a\in {\mathbb R}$ is a free parameter; the integration
contour ${\cal C}_\infty$ encompasses the point $z=\infty$; the
notation ${\bm t} \pm [{\bm z}^{-1}]$ stands for the infinite set of
parameters $\{t_j\pm z^{-j}/j\}$; for brevity, the physical
parameters ${\bm \varsigma}$ were dropped from the arguments of
$\tau$ functions.

Being expanded in terms of $\bm{ t^\prime}-{\bm t}$ and $a$,
Eq.~(\ref{bi-id}) generates four integrable hierarchies. One of
them, the Kadomtsev-Petviashvili (KP) hierarchy in the Hirota form
\cite{HirotaS} \vspace{-0.3cm}
\begin{equation}
   \label{kph}
    \frac{1}{2}\,D_1 D_k\, \tau_n^{(s)}(\bm{t})
    \circ \tau_n^{(s)}(\bm{t}) =  s_{k+1}([\bm{D}]) \, \tau_n^{(s)}(\bm{t})
    \circ \tau_n^{(s)}(\bm{t})
\end{equation}
($k\ge 3$) is of primary importance for the exact theory of replicas
\cite{Rem-mKP}. The first nontrivial member of the KP hierarchy
reads
\begin{eqnarray}
    \label{fkp}
    \left(
    \partial_{t_1}^4 + 3\,\partial_{t_2}^2 -
        4\, \partial_{t_1} \partial_{t_3}
    \right)\, \log \tau_n^{(s)}(\bm{\varsigma};{\bm t}) \nonumber \\
    + \,6\, \left(
        \partial_{t_1}^2\, \log \tau_n^{(s)}(\bm{\varsigma};{\bm t})
    \right)^2 = 0.
\end{eqnarray}
In what follows, it will be shown that its projection onto\linebreak
$s=0$ and ${\bm t}={\bm 0}$ [Eq.~(\ref{proj})] gives rise to a
nonlinear differential equation for the replica partition function
${\tilde Z}_n^{\rm(f/b)}({\bm {\varsigma}})$.

(ii) Since we are interested in deriving a differential equation for
${\tilde Z}_n^{\rm(f/b)}({\bm {\varsigma}})$ in terms of the
derivatives over {\it physical parameters} $\{\varsigma_j\}$, we
have to seek an additional block of the theory that would make a
link between the $\{t_j\}$ derivatives in Eq.~(\ref{fkp}) taken at
${\bm t}={\bm 0}$ and the derivatives over physical parameters
$\{\varsigma_j\}$. The study \cite{ASvM-1995} by Adler, Shiota, and
van Moerbeke suggests that the missing block is the {\it Virasoro
constraints} which reflect the invariance of the $\tau$ function
[Eq.~(\ref{tau-f})] under a change of the integration variables. In
the present context, it is useful to demand the invariance under the
transformation
\begin{eqnarray}
\label{v-tr}
\lambda_j \rightarrow \mu_j + \epsilon \mu_j^{q+1}f(\mu_j) \prod_{k=1}^{\dim({\bm c}^\prime)} (\mu_j-c_k^\prime),
 \;\;\epsilon>0,
\end{eqnarray}
where ${\bm c}^\prime=\{c_1,\cdots,c_{2r}\}\setminus\{\pm \infty\}$.
The function $f(\lambda)$ is related to the confinement potential
$V_{n-s}(\lambda)$ through the parameterisation $dV_{n-s}/d\lambda =
-g(\lambda)/f(\lambda)$, where $g(\lambda)=\sum_{j=0}^\infty b_j
\lambda^j$ and $f(\lambda)=\sum_{j=0}^\infty a_j \lambda^j$ depend
on $n$ and $s$.

The transformation (\ref{v-tr}) induces the Virasoro-like
constraints that can be written in the additive form
\begin{equation} \vspace{-0.3cm}
\label{2-Vir}
    \left[ \hat{{\cal L}}_{q}^V({\bm t}) + \hat{{\cal L}}_q^{\Gamma}({\bm \varsigma};{\bm t})
     \right] \tau_n^{(s)}({\bm \varsigma};{\bm t})
    =0, \;\;\; q \ge -1.
\end{equation}
The operators $\hat{{\cal L}}_{q}^V({\bm t})$ and $\hat{{\cal
L}}_{q}^\Gamma({\bm t})$ are associated with the
$e^{-V_{n-s}(\lambda)}$ and the $\Gamma({\bm \varsigma};\lambda)$
parts of the integrand, respectively; also, $\hat{{\cal
L}}_{q}^{\Gamma=1}({\bm t})\equiv 0$. The first operator
$\hat{{\cal L}}_{q}^V({\bm t})$ can be expressed in terms of the
Virasoro operators \cite{MM-1990}
\begin{eqnarray}
    \label{vo}
    \hat{{\cal L}}_q({\bm t}) = \sum_{j=1}^\infty jt_j \,\partial_{t_{q+j}}
    +
    \sum_{j=0}^q \partial_{t_j}\partial_{t_{q-j}}, \;\;\; q\ge -1,
\end{eqnarray}
which depend solely on the symmetry of the replica field theory and
obey, for all $p,q\ge -1$, the Virasoro algebra $
    [\hat{{\cal L}}_p,\hat{{\cal L}}_q] = (p-q)\hat{{\cal
    L}}_{p+q}$. [Equation (\ref{vo}) assumes that
$\partial_{t_0}$ is identified with the multiplicity of the matrix
integral in Eq.~(\ref{tau-f}), $\partial_{t_0} \equiv n$].
Explicitly, it holds \cite{OK-2007} that
\begin{widetext} \vspace{-0.5cm}
\begin{eqnarray}
\label{vLv}
     \hat{{\cal L}}_{q}^V({\bm t}) =
    \sum_{k=0}^{\dim ({\bm c}^\prime)} s_{{\dim ({\bm c}^\prime})-k}(-[{\bm \sigma}])
       \sum_{\ell = 0}^\infty \left(
        a_\ell \hat{\cal L}_{q+k+\ell}({\bm t}) - b_\ell \partial_{t_{q+k+\ell+1}}
    \right),\;\;\; [{\bm \sigma}]_j = \frac{1}{j}\sum_{k=1}^{\dim({\bm c}^\prime)} (c_k^\prime)^{j}, \;\;\;
    q\ge -1.
\end{eqnarray}
\end{widetext}
Here, $s_k({\bm t})$ are the Schur polynomials \cite{MD-1998}.

While very similar in spirit, the calculation of $\hat{{\cal
L}}_{q}^\Gamma({\bm t})$, the second ingredient in
Eq.~(\ref{2-Vir}), is more of an art since the function $\Gamma({\bm
\varsigma};\lambda)$ in Eq.~(\ref{tau-f}) may significantly vary
from one replica model to the other.

Remarkably, for ${\boldsymbol{t}={\boldsymbol 0}}$, the two
equations [Eqs. (\ref{fkp}) and (\ref{2-Vir})] can be solved jointly
to bring a closed nonlinear differential equation for ${\tilde
Z}_n^{(\rm{f/b})}({\boldsymbol {\varsigma}})$. It is this equation
which, being supplemented by appropriate boundary conditions,
provides a truly nonperturbative description of the replica
partition functions and facilitates performing the replica limit.

{\it Chiral GUE and bosonic replicas.}---To see the above formalism
at work and also answer the question raised in the title of our
Letter, let us consider the $N\times N$ chGUE random matrices
\begin{eqnarray} \vspace{-0.6cm}
\label{dm}
    \mathcal{ H}_{\mathcal D} = \left(
                 \begin{array}{cc}
                   0 & {\cal W} \\
                   {\cal W}^\dagger & 0 \\
                 \end{array}
               \right)
\end{eqnarray}
known to describe the low-energy sector of ${\rm SU}(N_c\ge 3)$ QCD
in the fundamental representation \cite{VZ-1993}. Composed of
rectangular $n_L\times n_R$ random matrices ${\cal W}$ with the
Gaussian distributed complex-valued entries
\begin{eqnarray}
\label{dm-den}
    P_{n_L,n_R}({\cal W}) =\left(
     \frac{2\pi}{N\Sigma^2}
    \right)^{n_L n_R}
     \exp\left[-\frac{N\Sigma^2}{2}{\rm tr\,} {\mathcal W}^\dagger {\mathcal W}\right],
\end{eqnarray}
where $N=n_L + n_R$, the matrix $\mathcal{ H}_{\mathcal D}$ has
exactly\linebreak $\nu =|n_R-n_L|$ zero eigenvalues identified with
the topological charge $\nu$; the remaining eigenvalues occur in
pairs $\{\pm \lambda_j\}$; the parameter $\Sigma$ denotes the chiral
condensate.

To determine the (microscopic) spectral density from the bosonic
replicas, we define the replica partition function
$Z_{n}^{\rm{(b)}}(\varsigma) = \left< {\rm det}^{-n}(\varsigma +
i\mathcal{H}_{\mathcal D})\right>_{\mathcal W}$ and map it onto a
bosonic field theory. In the half-plane ${\rm Re\,}\varsigma >0$,
the partition function $Z_{n}^{\rm{(b)}}(\varsigma)$ reduces to
\cite{DV-2001,F-2002}
\begin{equation}
\label{brft}
    {\tilde Z}_n^{\rm{ (b)}}(\omega) = \int_{{\mathcal S}_n}
    d\mu_{n}(\boldsymbol{\mathcal{Q}})\, \det{}^{\nu-n} \boldsymbol{\mathcal{Q}} \, \exp\left[
    -\frac{\omega}{2} {\rm Tr} (\boldsymbol{\mathcal{Q}} + \boldsymbol{\mathcal{Q}}^{-1})
    \right],
\end{equation}
where the integration domain ${\mathcal S}_n$ spans all $n\times n$
positive definite Hermitean matrices $\boldsymbol{\mathcal{Q}}$. Equation (\ref{brft}) was
derived in the thermodynamic limit $N\rightarrow \infty$ with the
spectral parameter $\omega=\varsigma N\Sigma$ being kept fixed (${\rm Re\,}\omega>0$).

Spotting the invariance of the integrand in Eq.~(\ref{brft}) under
the unitary rotation of the matrix $\boldsymbol{{\mathcal Q}}$, one
readily realises that ${\tilde Z}_n^{\rm{ (b)}}(\omega)$ belongs to
the class of $\tau$ functions specified by Eq.~(\ref{tau-f}) where
${\cal D}$ is set to ${\mathbb R}^+$, the potential $V_{n-s}$ is
$V_{n-s}(\lambda) = (n-s-\nu)\, \log \lambda$, and
$\Gamma({\boldsymbol\varsigma};\lambda)$ is replaced with $
\Gamma(\omega;\lambda) = \exp\left[-(\omega/2)
    (
            \lambda  + \lambda^{-1} )\right]
$. This observation implies that the associated $\tau$ function
$\tau_n^{(s)}(\omega; {\bm t})$ satisfies both the first KP equation
(\ref{fkp}) and the Virasoro constraints (\ref{2-Vir}) with
\cite{OK-2007} \vspace{-1.3cm}
\begin{widetext}
\begin{eqnarray}
\label{L-qv-bos}
   \hat{{\cal L}}_{q}^V({\bm t}) = \hat{{\cal L}}_{q+1}({\bm t}) + (\nu-n+s) \,\partial_{t_{q+1}},
   \;\;\; \hat{{\cal L}}_{q}^\Gamma(\omega;{\bm t}) =
    - \frac{\omega}{2} \partial_{t_{q+2}}
        -\delta_{q,\,-1}
  \left(
        \omega \partial_\omega + \frac{\omega}{2}\,\partial_{t_1}
    \right) +
    \left[1-\delta_{q,\,-1}\right]\, \frac{\omega}{2}\, \partial_{t_q}.
\end{eqnarray}
Projecting Eq.~(\ref{fkp}) taken at $s=0$ onto ${\bm t}={\bm 0}$,
and expressing the partial derivatives therein via the derivatives
over $\omega$ with the help of Eqs.~(\ref{2-Vir}) and
(\ref{L-qv-bos}), we conclude that ${\tilde Z}_n^{{\rm
(b)}}(\omega)=n! \,\tau_n^{(0)}(\omega;{\bm 0})$ obeys the
differential equation \cite{OK-2007}
\begin{eqnarray}
\label{fin-eq}
  h_n^{\prime\prime\prime} + \frac{2}{\omega} h_n^{\prime\prime}
  - \left( 4 + \frac{1+4(n^2+\nu^2)}{\omega^2}\right) h_n^{\prime}
  + 6(h_n^{\prime})^2
  + \frac{1 - 4 (n^2+\nu^2)}{\omega^3} h_n
  -\frac{2}{\omega^2} (h_n)^2 + \frac{4}{\omega} h_n h_n^{\prime}
  +\frac{4n^2}{\omega^2} = 0
\end{eqnarray}
\end{widetext}
that can be reduced to the Painlev\'e III. Here\linebreak
$h_n(\omega) =
\partial_\omega \, \log {\tilde Z}_n^{{\rm (b)}}(\omega)$.
Considered together with the boundary conditions
$h_n(\omega\rightarrow 0) \simeq -n\nu/\omega$ and\linebreak
$h_n(\omega\rightarrow \infty) \simeq -n - n^2/(2\omega)$, following
from Eq.~(\ref{tau-f}), the nonlinear differential equation
(\ref{fin-eq}) provides a nonperturbative characterisation of
${\tilde Z}_n^{\rm (b)}(\omega)$ for all $n\in {\mathbb Z}^+$.

To pave the way for the replica calculation of the Green function
$G(\omega)$ determined by the replica limit $G(\omega)=-\lim_{n
\rightarrow 0}n^{-1} h_n(\omega)$, one has to analytically continue
$h_n(\omega)$ away from $n$ integers. The previous studies
\cite{K-2002,K-2005} suggest that the sought analytic continuation
is given by the very same Eq.~(\ref{fin-eq}) where the replica
parameter $n$ is let to explore the entire real axis. This leap
makes the rest of the calculation straightforward. Representing
$h_n(\omega)$ in the vicinity of $n=0$ as\linebreak $h_n(\omega) =
\sum_{p=1}^\infty n^p a_p(\omega)$, we conclude that $G(\omega) =
-a_1(\omega)$ satisfies the equation
\begin{equation}
    \omega^3 G^{\prime\prime\prime} + 2 \omega^2 G^{\prime\prime}
  - \left( 1 + 4 \nu^2 + 4 \omega^2\right)\omega G^{\prime}
  + (1 - 4 \nu^2) \,G = 0.
\end{equation}
Its solution, subject to the boundary conditions consistent with
those specified below Eq.~(\ref{fin-eq}), brings the microscopic
spectral density $\varrho(\omega) = \pi^{-1} {\rm Re\,}
G(i\omega+0)$ in the form
\begin{eqnarray}
\label{den-fin}
    \varrho(\omega) = \nu \delta(\omega) + \frac{\omega}{2} \Big[
        J_\nu^2(\omega) - J_{\nu-1}(\omega) J_{\nu+1}(\omega)
    \Big].
\end{eqnarray}

Obtained within the framework of {\it bosonic} replicas, this
celebrated formula provides strong evidence against the idea of
their inapplicability to the nonperturbative description of random
matrix spectra \cite{VZ-1985}, in general, and of the chGUE spectra
\cite{DV-2001}, in particular. In view of the previous study
\cite{K-2002} on the performance of {\it fermionic} replicas, we are
led to speculate that truly nonperturbative approaches to nonlinear
replica $\sigma$ models leave no room for the fermionic-bosonic
dichotomy.

{\it Acknowledgements.} This work was supported by the Israel
Science Foundation through the grant No 286/04. 


\begin{references}
\bibitem{Rem-phys} This may impose some restrictions on the energies ${\boldsymbol \varsigma}$.
\bibitem{Int-D} Notice that ${\mathcal D}=[-1,+1]$ for (compact) fermionic replicas, and
    ${\mathcal D}=[0,+\infty]$ for (noncompact) bosonic replicas.
    A more general setting ${\cal D}=\bigcup_{j=1}^r [c_{2j-1},c_{2j}]$ does not complicate
    the theory we present.
\end{references}
\end{document}